\documentclass[a4paper,11pt]{article}
\usepackage{pos}
\usepackage{comment}

\makeatletter
\def\@preprint{\@empty}
\newcommand{\preprint}[1]{\gdef\@preprint{\hfill #1}}
\makeatother

\preprint{MIT-CTP/6170} 

\title{Unraveling Jet-Medium Interactions with Energy Correlators in Relativistic Oxygen Collisions}
\ShortTitle{Unraveling Jet-Medium Interaction with EECs in OO Collisions}

\author[a]{Arjun Srinivasan Kudinoor}
\author*[b]{Arthur Yi-Ting Lin}
\author[c,d]{Daniel Pablos}
\author[a]{Krishna Rajagopal}

\affiliation[a]{
  Center for Theoretical Physics --- a Leinweber Institute, Massachusetts Institute of Technology, Cambridge, MA 02139}

\affiliation[b]{
 Massachusetts Institute of Technology, Cambridge, MA, 02139
}

\affiliation[c]{Departamento de F\'isica, Universidad de Oviedo, Avda. Federico Garc\'ia Lorca 18, 33007 Oviedo, Spain}

\affiliation[d]{Instituto Universitario de Ciencias y Tecnolog\'ias Espaciales de Asturias (ICTEA), Calle de la Independencia 13, 33004 Oviedo, Spain}

\emailAdd{kudinoor@mit.edu}
\emailAdd{arthur72@mit.edu}
\emailAdd{pablosdaniel@uniovi.es}
\emailAdd{krishna@mit.edu}

\abstract{
As jet partons traverse a droplet of quark-gluon plasma (QGP) formed in a light- or heavy-ion collision, they can trigger hard scatterings off QGP quasiparticles. These hard scatterings broaden the hard and semi-hard core of jets, enhancing large-angle energy-energy correlations between jet particles. Oxygen-oxygen (OO) collisions provide an especially compelling arena to detect and study these hard scatterings since small QGP droplets in OO collisions reduce energy loss and selection bias. In this contribution to these proceedings, we present Hybrid Model calculations of energy-energy correlators (EECs) for $R=0.8$ inclusive jets in OO collisions with different $p_T$. We show that EECs of $R=0.8$ jets calculated using hadrons with $p_T>2$ GeV exhibit a bump at $R_L\sim0.3$--$0.5$. The bump angle corresponds to the typical angle of a hard $2 \rightarrow 2$ ``Moli\`ere'' scattering, and decreases as the jet $p_T$ increases. We further show semi-analytical pQCD calculations of the EEC for the two outgoing partons produced in a Moli\`ere scattering. This model-independent result qualitatively reproduces the $p_T$-dependence of the bump observed in Hybrid Model calculations, and shows how the angle of the left edge of the bump scales with the QGP temperature and the momentum-transfer threshold for Moli\`ere scattering in the model. Observing the bump and its dependence on jet $p_T$ in experiments would provide solid, model-independent, evidence that jet partons can resolve and probe the particulate structure of QGP.}

\FullConference{13th International Conference on Hard and Electromagnetic Probes of High-Energy Nuclear Collisions (HardProbes2026)\\
21-26 June 2026\\
Nashville, Tennessee, USA\\}

\begin{document}

\makeatletter
\noindent\@preprint
\makeatother

\maketitle

\begin{figure}
\centering
\includegraphics[width=0.495\linewidth]{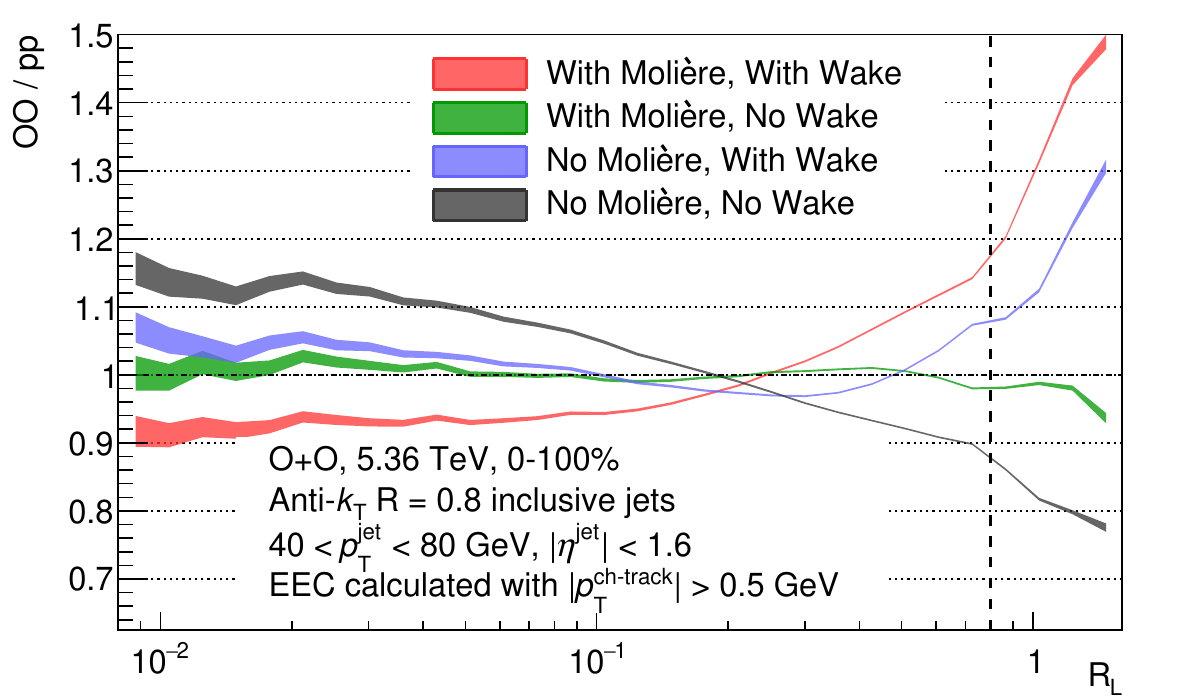}
\includegraphics[width=0.495\linewidth]{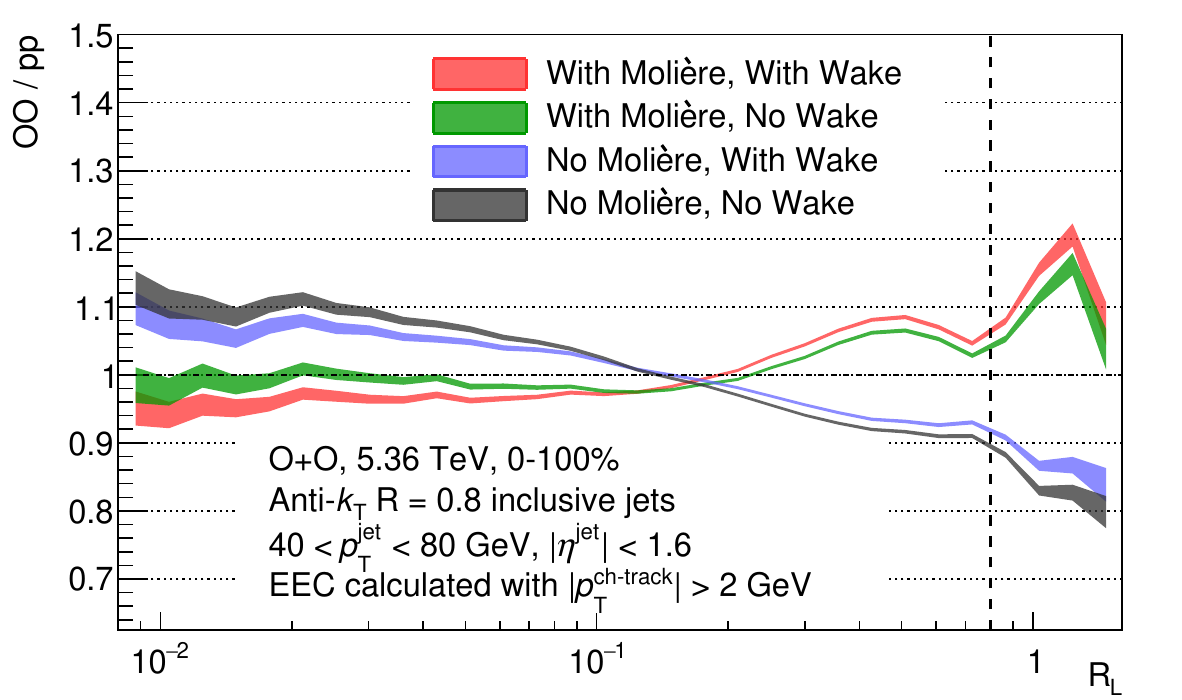} 
\caption{Hybrid Model calculations of the OO/pp ratio of jet EECs for $R=0.8$ jets with $40 \, {\rm GeV} < p^{\rm jet}_T < 80 \, {\rm GeV}$, calculated using charged particle tracks that satisfy $p_T > 0.5\, {\rm GeV}$ (left) and $p_T > 2\, {\rm GeV}$ (right). The dashed vertical line corresponds to $R_L = R_{\rm jet} = 0.8$.}
\vspace{-0.1in}
\label{fig:EEC_OO_4080}
\end{figure} 

As jet partons traverse a droplet of QGP produced in a light- or heavy-ion collision,
they lose energy to, and excite hydrodynamic wakes in, the strongly coupled medium. Energetic jet partons can also scatter at large angles  with high momentum transfer off  quark- and gluon-like quasiparticles in the droplet of QGP. Since droplets of QGP formed in oxygen-oxygen (OO) collisions are smaller than those formed in lead-lead collisions, the effects of jet energy loss and the resulting jet-selection bias on jet observables are smaller too~\cite{Kudinoor:2026wcs,Kudinoor:2026xxx}, but the internal structure of jets is still significantly modified by hard jet-medium interactions~\cite{Kudinoor:2026wcs}. 
The energy-energy correlator (EEC) observable can be used to study how the internal structure of jets is modified by their passage through QGP, making it a useful probe of hard scattering between jet partons and medium quasiparticles~\cite{Kudinoor:2026wcs}. 

\paragraph{The Hybrid Model.} The hybrid strong/weak coupling model (aka the Hybrid Model) is a theoretical framework for jet quenching in heavy- and light-ion collisions~\cite{Casalderrey-Solana:2014bpa}.
In the Hybrid Model, the production of high-$p_T$ partons and their subsequent showering according to the DGLAP evolution equations are described as implemented in PYTHIA 8. Each (light-like) parton in the shower loses energy and momentum as prescribed by a holographic energy loss formula~\cite{Chesler:2014jva,Chesler:2015nqz} as appropriate at strong coupling, as  described in Refs.~\cite{Casalderrey-Solana:2014bpa, Casalderrey-Solana:2016jvj,Casalderrey-Solana:2018wrw,Casalderrey-Solana:2019ubu}.
The lost energy is deposited back into the medium and excites hydrodynamic wakes~\cite{Casalderrey-Solana:2016jvj, Kudinoor:2025ilx, Bossi:2024qho}. In the Hybrid Model, the soft hadrons originating from the wake are sampled from a distribution given by a suitably modified Cooper-Frye prescription~\cite{Casalderrey-Solana:2016jvj}.

Hard scattering of jet partons off QGP quasiparticles is an additional, intrinsically weakly coupled, channel of jet-medium interactions. The Hybrid Model includes only the simplest hard jet-medium scattering process: large-angle, high momentum-transfer, $2\to 2$, Moli\`{e}re scattering between a jet parton and a QGP quasiparticle 
from the medium, assumed to be in thermal equilibrium,
with scattering probabilities calculated using tree-level pQCD amplitudes with QCD coupling $g_s=2.25$ meaning $\alpha_s \simeq 0.4$, as in Refs.~\cite{Hulcher:2022kmn, Hulcher:2026dht}.
Since soft jet-medium interactions 
are incorporated via strongly coupled energy loss, we ensure that the added elastic scattering is hard enough to be treated at weak coupling by requiring the momentum transfer to be above some threshold,
\begin{align}
|t|, |u| > a m_D^2, 
\end{align}
where $t$ and $u$ are Mandelstam variables and $m_D^2 = \frac{1}{3} g_s^2 T^2( N_c + N_f/2) =\frac{3}{2}g_s^2 T^2$ is the Debye mass squared. The threshold parameter $a$ in the Hybrid Model is set to $10$ in all calculations below. 

\paragraph{EECs in OO Collisions.}

\begin{figure}
\centering
\begin{minipage}{0.45\textwidth}
\caption{Hybrid Model calculations of OO/pp ratios of EECs with (red) and without (blue) Moli\`ere scattering for $R_{\rm jet} = 0.8$ jets with $p^{\rm jet}_T \in [40, 80]$, $[80, 120]$, and $[120, 160]$ GeV, calculated using charged particles with $p_T>2$ GeV. Darker shades correspond to higher $p_T^{\rm jet}$. The dashed vertical line corresponds to $R_L = R_{\rm jet}$; the dotted vertical lines in different shades of red are drawn at the positions where the OO/pp ratios of EEC($R_L$) attain their respective local maxima.} \label{fig:EEC_vary_pT}
\end{minipage}
\begin{minipage}{0.54\textwidth}
\centering 
\includegraphics[width=0.99\textwidth]{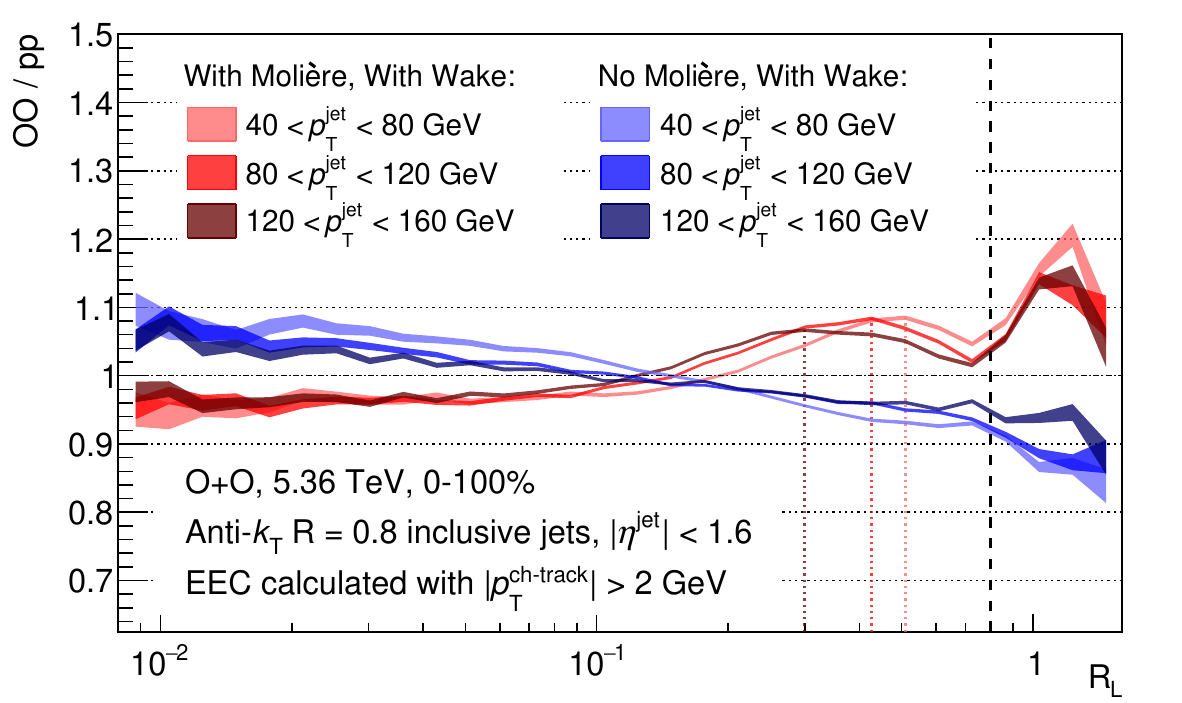} 
\end{minipage}
\vspace{-0.1in}
\end{figure} 

We present Hybrid Model calculations of EECs in OO collisions. We follow the jet reconstruction procedure and definition of the EEC observable from Ref.~\cite{CMS:2025ydi}:  
\begin{align}
\mathrm{EEC}(R_L) = \frac{1}{\mathcal{N}} \frac{1}{\delta r} \sum_{{\rm jets}} \sum_{(i,j)\in {\rm   pairs}} p_T^{i} p_T^{j} \Big|_{\Delta R_{ij} \in \left[R_L - \frac{\delta r}{2} ,R_L + \frac{\delta r}{2} \right] }, 
\end{align}
where the particle pairs are charged hadrons that are separated from the jet axis by $\Delta R < R_{\rm jet}$ and where $\mathcal{N}$ is a normalization factor chosen such that $\int_{0.008}^{R_{\rm jet}} \mathrm{d} R_L\,  \mathrm{EEC}(R_L) = 1$.

Fig.~\ref{fig:EEC_OO_4080} shows Hybrid Model calculations of OO/pp ratios of $\mathrm{EEC}(R_L)$ for $R=0.8$ inclusive jets, from Ref.~\cite{Kudinoor:2026wcs}. It is helpful to consider three different regions of angular separation $R_L$: (i) small-angle region $R_L \lesssim 0.2$; (ii) large-angle region, $0.2 \lesssim  R_L < R_{\rm jet}$; (iii) very-large-angle region, $R_{\rm jet} < R_L \lesssim 2R_{\rm jet}$. The small-angle region is
dominated by correlations of particles within the hard jet core. In the absence of Moli\`{e}re scattering (grey and blue bands), the OO/pp ratio is above one at small $R_L$. This is due to jet-selection bias: since the number of jets produced in a collision falls steeply as a function of $p_T$, the sample of jets in OO collisions selected within a given $p_T$-bin are biased towards those that lost the least amount of energy to the plasma; these jets tend to have narrower hard cores in the absence of hard jet-medium scattering effects.
However, when Moli\`{e}re scattering is included (green and red color bands), the OO/pp ratio can be below one at small $R_L$, reflecting a genuine broadening of the hard jet core due to hard jet-medium interactions.

The large-angle region $0.2 \lesssim R_L < R_{\rm jet}$ is dominated by correlations of particle-pairs where one particle is in the hard jet core and the other is in the periphery of the jet. In the left panel of Fig.~\ref{fig:EEC_OO_4080}, including either the wake (blue) or Moli\`{e}re scattering (green) alone enhances the OO/pp ratio at large $R_L$; when both effects are included (red), the OO/pp ratio is enhanced even further. 
If we groom away  soft hadrons 
(including those from jet wakes)
by imposing a track cut $p_T>2$ GeV (right panel), the OO/pp ratio of EEC($R_L$) remains above unity
--- a consequence of hard jet-medium interactions. Furthermore, 
this ratio
exhibits a local maximum --- a bump --- at an angle $R_L \simeq 0.5$ only when Moli\`{e}re scattering is included (red and green). Intuitively, the position of the bump corresponds to a typical angle of Moli\`{e}re scattering. Fig.~\ref{fig:EEC_vary_pT} confirms this interpretation by showing the jet-$p_T$ dependence of the OO/pp ratios of EEC$(R_L)$ for $R_{\rm jet}=0.8$ jets with $p_T^{\rm ch-track} > 2$ GeV. Since higher momentum partons are more likely to scatter at smaller angles~\cite{Hulcher:2026dht}, the $R_L$-position of this bump is expected to decrease with increasing jet $p_T$, as is seen in the red bands.
Note that this bump can only be seen for jets with radii greater than the typical angle of a Moli\`ere scattering.

Finally, the very-large-angle region probes correlations between pairs of particles where neither  belongs to the central core of the jet. In the left panel of Fig.~\ref{fig:EEC_OO_4080}, we see that the OO/pp ratios of EEC($R_L$) are enhanced primarily because soft hadrons from jet wakes populate the periphery of the jet. From the very-large-angle region in the right panel of Fig.~\ref{fig:EEC_OO_4080}, we see that partons from  Moli\`{e}re scatterings also populate the periphery of jets, in this case with hadrons having $p_T^{\rm ch-track}>2$ GeV.

\begin{figure}
    \centering
     \includegraphics[width=0.329\linewidth]{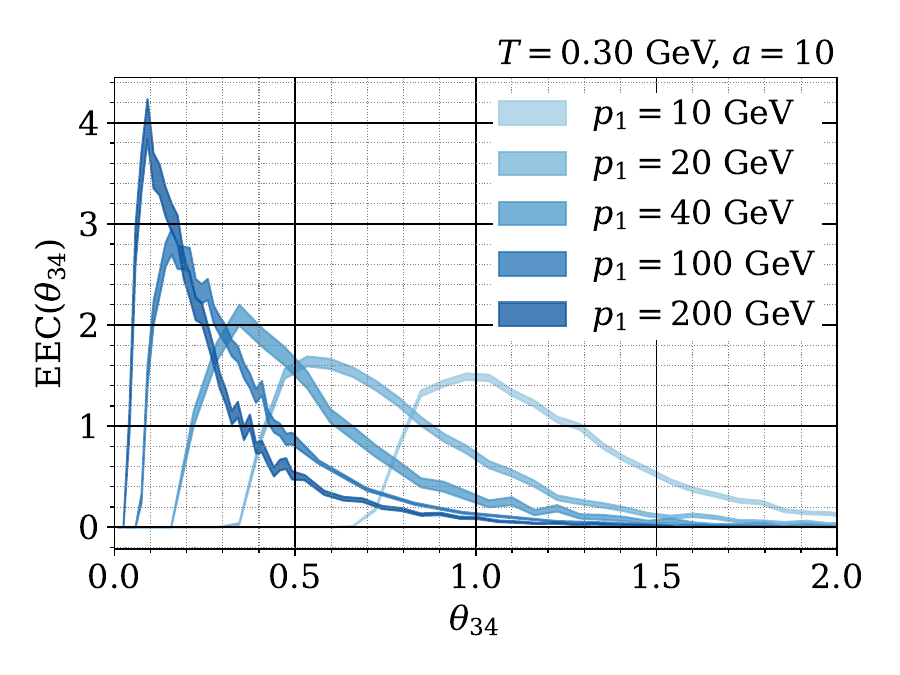}
    \includegraphics[width=0.329\linewidth]{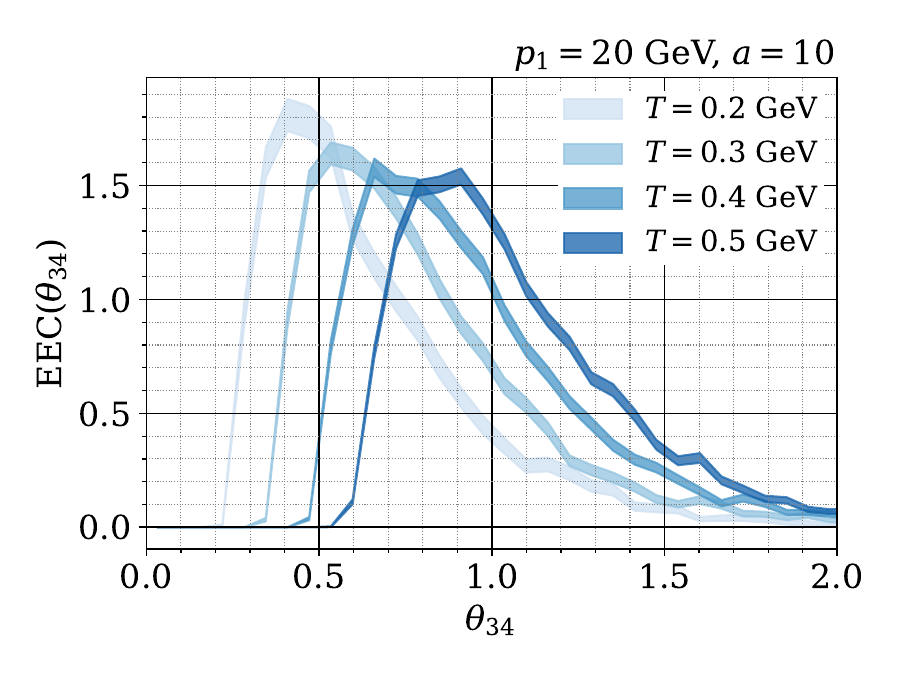}
    \includegraphics[width=0.329\linewidth]{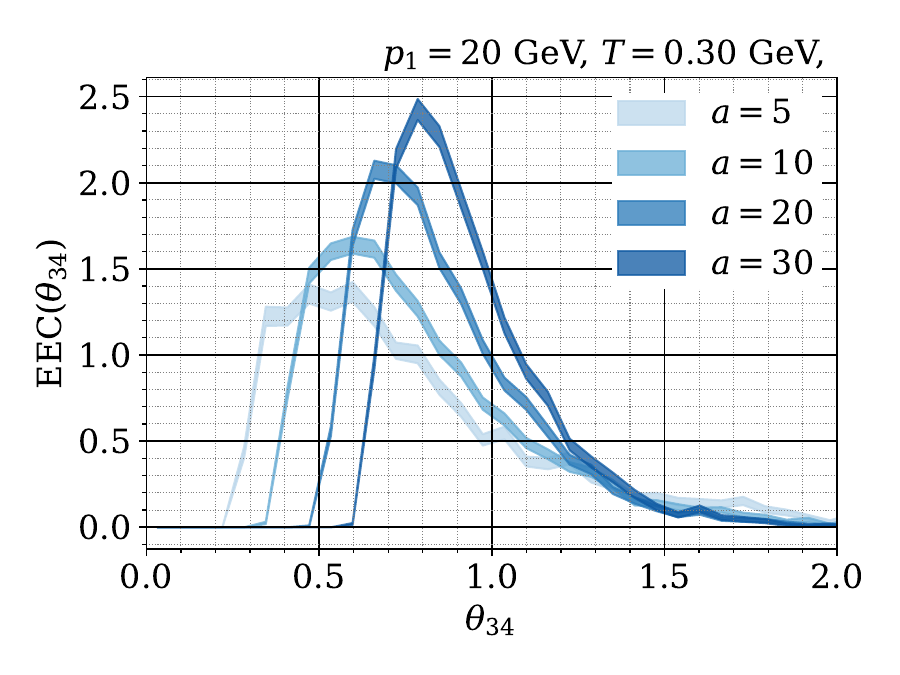} 
    \includegraphics[width=0.329\linewidth]{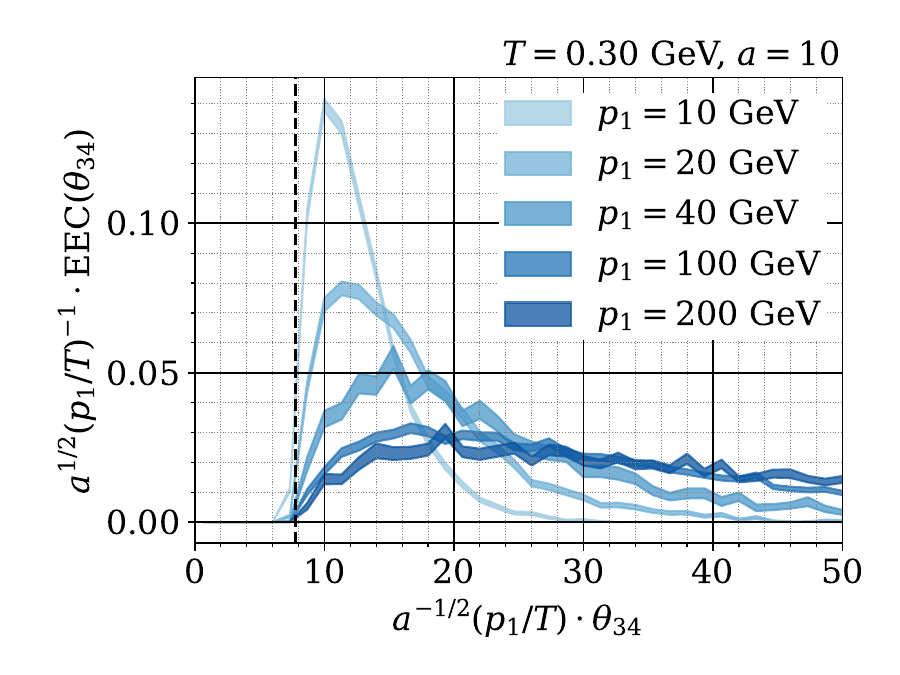}
    \includegraphics[width=0.329\linewidth]{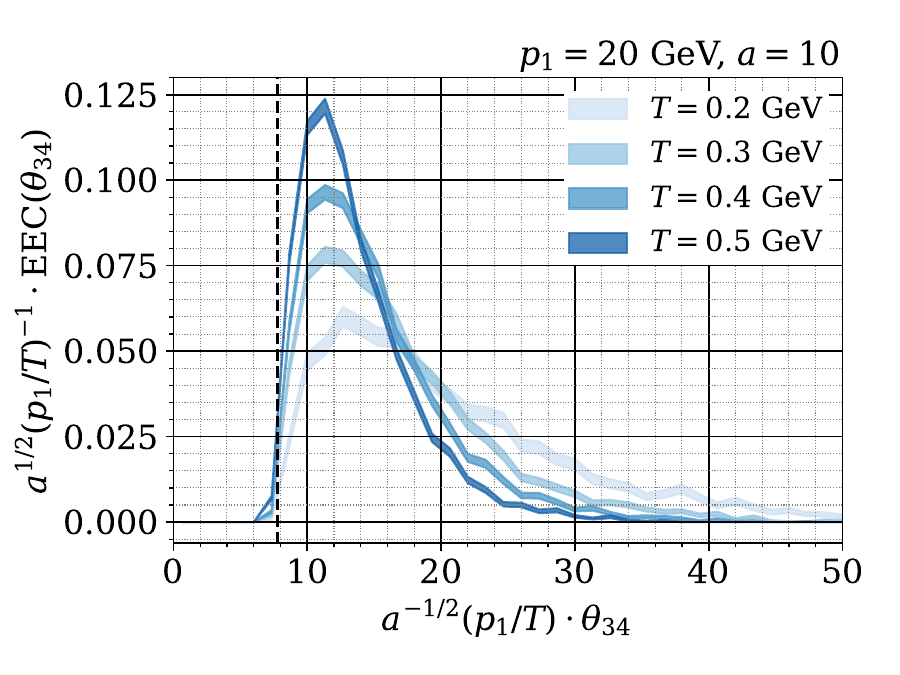} \includegraphics[width=0.329\linewidth]{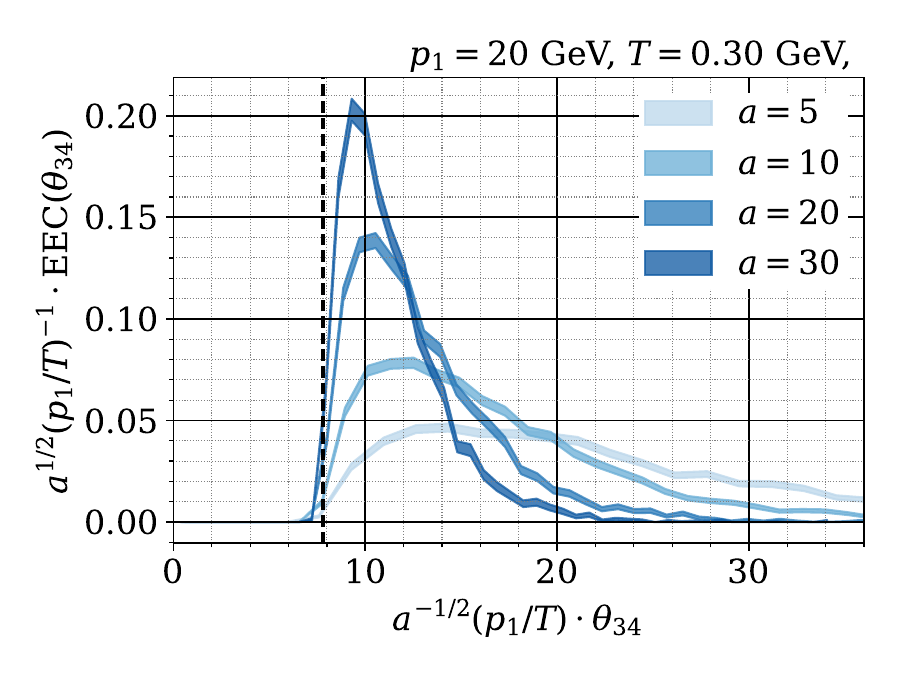}
    \caption{Self-normalized EEC of two outgoing particles from a Moli\`{e}re scattering. EECs are are calculated while varying the jet parton momentum $p_1$ (left), varying the medium temperature $T$ (center), and varying the momentum-transfer threshold parameter $a$  (right). The horizontal axes of the upper panels are the opening angle $\theta_{34}$ and the horizontal axes of the lower panels are $a^{-1/2}(p_1/T) \cdot \theta_{34}$, rescaled to align the left edge of the EEC distributions. The black dashed lines on the lower panels are drawn at $\theta_{34} = 2 \sqrt{2 a} \, m_D/p_1$.}
    \vspace{-0.1in}
    \label{fig:analytic_EEC}
\end{figure}

\paragraph{pQCD Calculations of Moli\`ere EECs.} We now investigate the bump in the OO/pp ratio of EECs, first discovered via Hybrid Model calculations~\cite{Kudinoor:2026wcs},
using model-independent pQCD calculations. To do so, we calculate the EEC of two outgoing particles that result from a single Moli\`{e}re scattering between a jet parton (1) and a QGP quasiparticle (2), as a function of the angle $\chi = \theta_{34}$ between the outgoing partons (3 and 4):
\begin{align}
\frac{\mathrm{d} \Sigma}{\mathrm{d} \chi} = \frac{1}{\sigma} \int \mathrm{d} \sigma   \frac{|\boldsymbol{p}_3| |\boldsymbol{p}_4|}{|\boldsymbol{p}_1|^2}\delta(\chi -\theta_{34}). 
\end{align}
The incident jet parton has fixed momentum $\boldsymbol{p}_1$, while the incoming medium parton has momentum $\boldsymbol{p}_2$, sampled from a Boltzmann distribution with fixed temperature $T$. We require that each scattering satisfy $|t|,|u|>am_D^2$ as in the Hybrid Model calculations, but now treat the threshold parameter $a$ as a variable which is not fixed {\it a priori}. 
In the regime $a m_D^2 \lesssim |t| \ll s$, the leading-order pQCD matrix elements for elastic $2 \rightarrow 2$ partonic scattering reduce to a Rutherford-like form $|\mathcal{M}_{ab\to ab}|^2 = C_{ab} g_s^4 (s^2+u^2)/t^2$, where $C_{ab} = 4/9$, $1$, and $9/4$ for quark-quark, quark-gluon, and gluon-gluon scattering. For simplicity, we adopt this form throughout the calculations below. 

The upper panels of Fig.~\ref{fig:analytic_EEC} show self-normalized EEC distributions of the two outgoing particles sampled from single Moli\`{e}re scattering events, for different values of the incident parton momentum $p_1$ (upper left), the medium temperature $T$ (upper center), and momentum-transfer threshold parameter $a$ (upper right). Each EEC distribution has a maximum in the range $0.1 \lesssim \theta_{34} \lesssim 1$ which exhibits similar systematic behavior as the bump seen in Hybrid Model calculations of the OO/pp ratios of EECs: the angle $\theta_{34}$ where the EEC for Moli\`ere scattering peaks decreases with increasing incident parton momentum. The peak value of $\theta_{34}$ also increases as $T$ or $a$ increases.

The dependence of the peak value of $\theta_{34}$ on $p_1$, $T$, and $a$ can be understood from the kinematics of the $2\to2$ scattering. For massless partons, $s=2|\boldsymbol{p}_3||\boldsymbol{p}_4|(1-\cos\theta_{34})$. In the limit $p_1\gg T$ and for small $\theta_{34}$, this gives $s\simeq z(1-z)p_1^2\theta_{34}^2$, where $z$ is the fraction of the outgoing pair's energy carried by either one of the partons. Equivalently, the angle is given by $\theta_{34} \simeq \sqrt{s/z(1-z)}/p_1$. Since $z(1-z) \leq 1/4$, we find $\theta_{34} \gtrsim 2 \sqrt{s}/p_1$. 
Furthermore, the requirement $|t|,|u|>am_D^2$ 
imposes $s> s_{\rm min}=2am_D^2$ which, together with $z(1-z)\leq 1/4$, implies
a minimum allowed $\theta_{34}$:
\begin{equation}
    \theta_{34}>\theta_{34}^{\rm min} \approx 2\sqrt{2a} \, m_D/p_1 
    = 2\sqrt{3a} \, g_s\,T/p_1.
\label{eq:theta34minscaling}
\end{equation}
The fact that there is a minimum allowed $\theta_{34}$ is apparent in Fig.~\ref{fig:analytic_EEC}.
To confirm that $\theta^{\rm min}_{34}$ scales as in Eq.~\eqref{eq:theta34minscaling}, 
in the lower panels of Fig.~\ref{fig:analytic_EEC}
we plot the EEC (rescaled so as to keep it self-normalized)
against the rescaled angle $a^{-1/2}(p_1/T)\theta_{34}$. 
As expected, the left edges of the distributions collapse onto the same value, $a^{-1/2}(p_1/T)\theta_{34}^{\rm min} \simeq
2\sqrt{3}\,g_s$, independent of $p_1$, $T$, and $a$. This confirms 
that 
in the model implementation~\cite{Hulcher:2022kmn,Hulcher:2026dht} of the
hard jet-medium elastic scattering process that we have employed,
the onset of the effects of Moli\`ere scattering 
on the EEC observable is indeed set by the momentum-transfer threshold, which implies $s>2a m_D^2$.

Inspection of the upper panels of Fig.~\ref{fig:analytic_EEC} indicates that the peak in the EEC occurs at an angle $\theta_{34}\sim (1.3-2.5)\theta_{34}^{\rm min}$.
It is straightforward to see that the peak in the EEC must correspond to scatterings with $s\ll 12 p_1 T$: reaching such a large value of $s$ requires an incident thermal parton that is either almost precisely back-to-back with the jet parton or that has a 
Boltzmann-suppressed momentum or both, which makes values of $s$ this large rare.
We have checked that in all cases plotted in Fig.~\ref{fig:analytic_EEC} the scatterings that populate the peak of the EEC distribution correspond to values of $s$ that are $\sim (2-4)s_{\rm min}$, with varying values of $z$ in different cases plotted.
This corresponds to $s\ll 12 p_1 T$ for the larger values of $p_1/T$ that we have considered, and $s< 12 p_1 T$ in all cases.

In a companion contribution to these proceedings~\cite{Kudinoor:2026xxx}, we demonstrate that both the large-angle EEC enhancement and its characteristic bump, discovered via Hybrid Model calculations~\cite{Kudinoor:2026wcs} and analyzed here via a model-independent pQCD calculation,
are very similar when the 
observable is constructed using only charged particles with $p_T>p_T^{\rm min}=4$~GeV. Here, as in Ref.~\cite{Kudinoor:2026wcs}, we have used $p_T^{\rm min}=2$~GeV,
which is sufficient to eliminate contributions from jet wakes in the Hybrid Model. In the 
Hybrid Model, though, the soft hadrons originating from jet wakes are somewhat too soft~\cite{Casalderrey-Solana:2016jvj}, making $p_T^{\rm min}=4$~GeV a safer choice in any model.

Experimental measurements of the enhancement of semi-hard large-angle correlations in EECs in OO collisions, the bump therein, and the jet-$p_T$-dependence of the angle where the bump is found, would provide direct,  exciting, evidence for hard scattering between jet partons and quasiparticles in QGP.
Hybrid Model calculations~\cite{Kudinoor:2026wcs,Kudinoor:2026xxx} also indicate that hard jet-medium scatterings enhance the population of jets with large Soft Drop angle $R_g$ in OO collisions. Experimental measurements of this enhancement would provide complementary evidence for hard jet-medium scattering.

\noindent
\paragraph{Acknowledgments.} Research supported in part by the U.S.~DOE, Office of Science, Office of Nuclear Physics grant
DE-SC0011090,
by Ram\'on y
Cajal fellowship RYC2023-044989-I,
by the NSF Graduate Research Fellowship Program
grant~2141064, 
and by subMIT at MIT Physics~\cite{Acosta:2025kgk}.

\bibliographystyle{JHEP}
\bibliography{bibliography}

\end{document}